\documentclass[aps,pra,10pt,twocolumn,superscriptaddress,floatfix]{revtex4-2}
\pdfoutput=1

\usepackage{amsmath}
\usepackage{amssymb}
\usepackage{amsthm}
\usepackage{mathrsfs}
\usepackage{bm, dsfont}
\usepackage[usenames,dvipsnames]{color}
\usepackage{colortbl}
\usepackage[table]{xcolor}
\usepackage{enumitem}
\usepackage{graphicx}
\usepackage{natbib}
\usepackage[colorlinks=true,linkcolor=blue,citecolor=blue,urlcolor=blue]{hyperref}
\usepackage{hypcap}
\usepackage{verbatim, float}
\usepackage{psfrag}
\usepackage[normalem]{ulem}
\usepackage{physics}

\newcommand{\cH}{\mathcal{H}}

\newcommand{\be}{\begin{equation}}
\newcommand{\ee}{\end{equation}}

\newtheorem*{observation}{Observation}
\newtheorem{theorem}{Theorem}
\newtheorem{corollary}[theorem]{Corollary}

\newcommand{\hA}{\mathcal{A}}
\newcommand{\hB}{\mathcal{B}}

\newcommand{\hL}{\mathcal{L}}

\newcommand{\hO}{\mathcal{O}} 

\newcommand{\hS}{\mathcal{S}}
\newcommand{\hT}{\mathcal{T}}

\newcommand{\N}{\mathbb N} 

\newcommand{\fii}{\varphi} 

\newcommand{\hi}{\mathcal{H}} 
\newcommand{\ki}{\mathcal{K}} 

\newcommand{\lh}{\mathcal{L(H)}} 
\renewcommand{\th}{\mathcal{T(H)}} 
\newcommand{\sh}{\mathcal{S(H)}} 

\def\<{\langle} 
\def\>{\rangle} 

\newcommand{\pX}{\sigma_{X}}
\newcommand{\pZ}{\sigma_{Z}}

\newcommand{\kb}[2]{|#1 \rangle\langle #2|} 

\newcommand{\linspan}{\mathrm{span}}
\newcommand{\tmax}{\hat{\otimes}}
\newcommand{\btn}[1]{$1 \mapsto #1$}

\renewcommand{\paragraph}[1]{\addcontentsline{toc}{section}{#1}\emph{#1.}---}

\begin{document}

\title{No-broadcasting characterizes operational contextuality}

\author{Pauli Jokinen}
\email{pauli.jokinen@etu.unige.ch}
\author{Mirjam Weilenmann}
\affiliation{Department of Applied Physics, University of Geneva, Switzerland}
\author{Martin Plávala}
\affiliation{Naturwissenschaftlich-Technische Fakult\"{a}t, Universit\"{a}t Siegen, Walter-Flex-Stra\ss e 3, 57068 Siegen, Germany}
\author{Juha-Pekka Pellonpää}
\affiliation{Department of Physics and Astronomy, University of Turku, FI-20014 Turun yliopisto, Finland}
\author{Jukka Kiukas}
\affiliation{Department of Mathematics, Aberystwyth University, Aberystwyth SY23 3BZ, United Kingdom}
\author{Roope Uola}
\affiliation{Department of Applied Physics, University of Geneva, Switzerland}

\begin{abstract}
Operational contextuality forms a rapidly developing subfield of quantum information theory. However, the characterization of the quantum mechanical entities that fuel the phenomenon has remained unknown with many partial results existing. Here, we present a resolution to this problem by connecting operational contextuality one-to-one with the no-broadcasting theorem. The connection works both on the level of full quantum theory and subtheories thereof. We demonstrate the connection in various relevant cases, showing especially that for quantum states the possibility of demonstrating contextuality is exactly characterized by non-commutativity, and for measurements this is done by a norm-1 property closely related to repeatability. Moreover, we show how techniques from broadcasting can be used to simplify known foundational results in contextuality.
\end{abstract}

\maketitle

\section{Introduction}
Contextuality is a fundamental notion that was originally introduced to rule out (deterministic) hidden variable models for quantum theory~\cite{KochenSpecker}.
It captures the idea that measurements in quantum theory generally cannot be considered as revealing pre-existing classical values independent of the measurement context. In contrast to Bell inequalities~\cite{Bell} that are concerned with space-like separated parties, contextuality can be defined in prepare-and-measure scenarios on a single system. Contextuality has been established as a feature of quantum theory that has been shown -- in a similar vein as the presence of entanglement -- to be a prerequisite for various quantum advantages, including in quantum computation~\cite{Raussendorf, Howard2014, Delfosse, BV1} and communication tasks~\cite{Saha, Spekkens2009, Schmid2018, Cubitt}.

While the original notion of contextuality was referring to projective quantum measurements, this has since been extended to a notion that applies to more general operational theories and unsharp measurements~\cite{Spekkens2005}. This broader form of contextuality includes two central concepts: 
\emph{preparation contextuality} and \emph{measurement contextuality}, allowing one to investigate more refined classical models. These have recently experienced a surge in research activity in both quantum and general operational theories~\cite{fragments}, and different types of contextuality have been further connected to other foundational concepts such as measurement incompatibility~\cite{TavakoliUola,plavala2022incompatibility,Selby2023}, entanglement~\cite{plavala2024contextuality}, steering~\cite{TavakoliUola,plavala2022incompatibility}, negativity~\cite{SpekkensNeg} and anomalous weak values~\cite{Pusey2014, Kunjwal2019}.

\begin{figure}
\centering
\includegraphics[width=\linewidth]{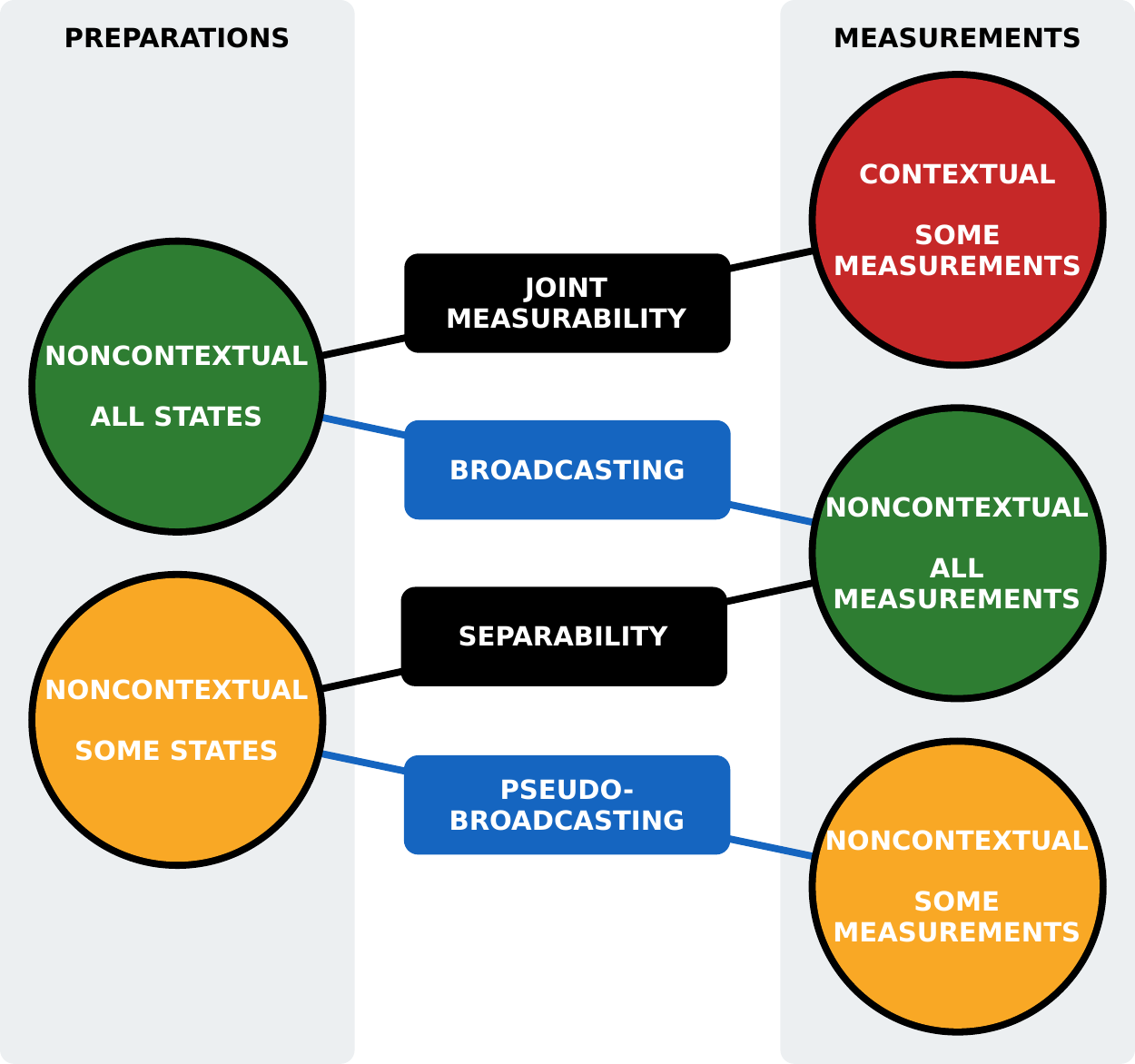}
\caption{Non-contextuality of an operational theory and the corresponding quantum mechanical entities. On the left (resp. right) are the properties requested from an ontological model for states (resp. measurements). The color green refers to ontological models for full theory, yellow refers to subtheories, and red refers to models exhibiting contextuality. The binding quantum mechanical entity is given in the middle. The black connections represent known results from Refs.~\cite{TavakoliUola,plavala2022incompatibility,Selby2023,plavala2024contextuality}. The contributions of this work are given by the blue connections.}
\label{fig:compare2}
\end{figure}

In this work, we ask whether there is a fundamental quantum entity that fully characterizes generalized contextuality. As our main result, we give a positive answer to this question by mapping contextuality one-to-one with the no-broadcasting theorem~\cite{NoBroadcasting, Barnum2006}. The proof of this fact is based on the realization that non-contextual models can be interpreted as measure-and-prepare mechanisms, where only classical information passes through. One direction of the proof follows since classical information is broadcastable, and the other is based on an averaging argument over broadcasting channels. This shows that generalized contextuality captures a well-established operational entity in quantum theory, and that advantages provided by contextuality can alternatively be attributed to the no-broadcasting theorem. We further show that for subtheories of quantum theory the characterization is given by pseudo-broadcasting.

We demonstrate the power of the connection by using results from broadcasting~\cite{NoBroadcasting, Barnum2006} in the realm of contextuality. First, we completely characterize the sets of quantum states and measurements that allow for proofs of contextuality of quantum theory by using the no-broadcasting theorem. This complements a line of research on characterizing and witnessing different facets of contextuality using entities of quantum theory \cite{liang11,yu13,Kunjwal14,kunjwal2014note,LostaglioWork,Xu19,TavakoliUola,plavala2022incompatibility,Selby2023,plavala2024contextuality}. In contrast, we characterize full, i.e. preparation and measurement, contextuality, cf.\ Fig.~\ref{fig:compare2}. This allows one to immediately obtain known results showing contextuality without incompatibility \cite{Selby2023} and robustness of contextuality to arbitrary amounts of noise \cite{Rossi2023}, as analogous results are known for broadcasting \cite{heinosaari2016simultaneous,heinosaari2019no,mitra2021layers,Heinosaari_2023}. Second, we show that the connection is beneficial for understanding the relationship between contextuality and other established notions of classicality given by joint measurability \cite{HeinosaariInvitation,GuhneJMReview}, non-disturbance \cite{Heinosaari10}, and Leggett-Garg macrorealism~\cite{Leggettoriginal,EmaryLGReview,VitaglianoBudroniReview,SchmidMacrorealism}. Finally, checking directly whether an arbitrary sub-theory of quantum theory allows for a non-contextual model is in general not a straightforward task. We demonstrate how pseudo-broadcasting can be used to derive non-contextuality criteria in such setting.

\section{Operational contextuality}
In measurement and preparation non-contextual theories, the probabilities of the theory can be reproduced in the following way. Let $\{P\}$ be the convex set of preparations and $\{M\}$ the convex set of measurements of the theory with outcomes labelled by $k$. Then the probabilities are reproduced according to the following \textit{ontological model}:
\begin{align}
    p(k|P,M)=\sum_{\lambda} \mu_P(\lambda) \xi_{M,k}(\lambda). \label{ontologicalmodel1}
\end{align}
Here $\mu_P(\lambda)$ are probabilities with $\sum_{\lambda} \mu_P(\lambda)=1$ for all $P$ and $\xi_{M,k}(\lambda)$ are indicator functions with $\sum_k \xi_{M,k}(\lambda)=1$ for all $M$ and $\lambda$. Furthermore $\mu_P$ and $\xi_{M,k}$ depend only on the operational equivalence classes \cite{Spekkens2005}, i.e., classes of preparations (resp. measurements) that are not distinguishable by any measurements (resp. states). This motivates the term \textit{operational contextuality} and implies that $\mu_P$ and $\xi_{M,k}$ are affine with respect to (equivalence classes of) preparations and measurements respectively \cite{muller2023testing}.

We now ask what this notion of non-contextuality means in quantum theory. In this case, preparation equivalence classes are formally described as quantum states, i.e.\ positive semi-definite unit-trace linear operators acting on a finite-dimensional Hilbert space, denoted by $\rho$. Furthermore, equivalence classes of measurements, or observables, are described by positive operator valued measures (POVMs), i.e.\ collections of positive linear operators $M=\{M_k\}_k$ with the normalization $\sum_k M_k=I$. In quantum theory probabilities are produced via the Born rule $p(k|\rho,M)=\tr{\rho M_k}$. Using basic Hilbert space duality and the affinity requirements of $\mu_{\rho}$ and $\xi_{M_k}$ one sees that Eq.~(\ref{ontologicalmodel1}) gets the form
\begin{align}
    \tr{\rho M_k}=\sum_{\lambda} \tr{\rho G_{\lambda}}\tr{\sigma_{\lambda}M_k},
    \label{ontologicalmodel3}
\end{align}
where $\{G_\lambda\}_\lambda$ and $\{\sigma_\lambda\}_\lambda$ are such that $\tr{\rho G_{\lambda}} \geq 0$, $\sum_\lambda \tr{\rho G_{\lambda}} = 1$, and $\tr{\sigma_{\lambda} M_k} \geq 0$ for all $\lambda$, all states $\varrho$ and all POVMs $\{M_k\}_k$. In other words, a non-contextual model for quantum theory is equivalent to Eq.~(\ref{ontologicalmodel3}) with $\{G_{\lambda}\}_\lambda$ being a POVM and $\{\sigma_{\lambda}\}_\lambda$ being quantum states.
 
\section{Broadcasting}
Classical information can be copied arbitrarily many times in an exact manner. In quantum theory the same idea is captured by the concept of broadcasting and the inability to copy information stored in a quantum state is known as the no-broadcasting theorem \cite{NoBroadcasting,Barnum2006}. To define broadcasting, we need the notion of \emph{quantum channels}. A quantum channel $\Lambda: \lh \to \hL(\ki)$ is a completely positive trace preserving linear map. Here $\hi$ and $\ki$ denote finite dimensional Hilbert spaces and $\lh$ is the set of linear operators acting on $\hi$. Let now $\hT$ be a subset of states and $\hA,\hB$ subsets of observables acting on $\hi$. Then the triple $(\hT,\hA,\hB)$ is called \emph{broadcastable} if there exists a channel $\Lambda:\lh \to \hL(\hi \otimes \hi)$ such that the following holds for all $\{A_k\}_k \in \hA$, $\{B_k\}_k \in \hB$, $\rho \in \hT$, and $\forall k$:
\begin{align}\label{kakkonen}
\tr{A_k \tr_2{\Lambda(\rho)}}&=\tr{A_k \rho}, \\
\tr{B_k \tr_1{\Lambda(\rho)}}&=\tr{B_k \rho}.
\label{kolmone}
\end{align}
Here $\tr_i(\cdot)$ for $i=1,2$ denotes the partial trace over subsystem $i$. One can directly extend this definition to a notion of \btn{n}\emph{-broadcasting}. The tuple $(\hT,\hA_1,\dots,\hA_n)$ is \btn{n}-broadcastable, if there exists a channel $\Lambda:\lh \to \hL(\hi^{\otimes n})$ such that for all $i \in \{1,\dots, n\}$ and for all $\{A_{i,k}\}_k \in \hA_i$ the condition $\tr{A_{i,k}\tr_{i^c} \Lambda(\rho)}=\tr{A_{i,k}\rho}$ holds for all $\rho \in \hT$. Here $\tr_{i^c}(\cdot)$ denotes the partial trace over the complement of $\{i\}$ i.e.\ the subsystems not labeled by the index $i$. Note that we may assume above that $\hT$ is convex, by passing to the convex hull if necessary, since \eqref{kakkonen} and \eqref{kolmone} are linear with respect to $\rho$.

\section{Connection between operational non-contextuality and broadcasting} Quantum theory is contextual. It is natural to what the operational ingredients that are responsible for this phenomenon are. In this section, we give an answer to this question. We first state the main observation, the validity of which follows from the two special cases presented in Theorem~\ref{thmstates} for states and Theorem~\ref{thmmeasurement} for measurements.

\begin{observation}
Any proof of contextuality of quantum theory is equivalent to a proof of no-broadcasting.
\end{observation}

It is apparent from Eq.~(\ref{ontologicalmodel3}), that a subset of states that does not allow for a proof of contextuality of quantum theory, has to be a subset of the fixed points of some measure-and-prepare or entanglement breaking channel (EBC) $\Lambda$ given by $\rho=\sum_{\lambda} \tr{\rho G_{\lambda}} \sigma_{\lambda}=:\Lambda(\rho)$ \cite{Horodecki03}. This formulation encodes the intuition of non-contextual models being measure-and-prepare models allowing only classical information $\lambda$ to pass through from the measurement $\{G_\lambda\}_\lambda$ to the preparation $\{\sigma_\lambda\}_\lambda$. Using this, we can prove the following theorem. Here $\mathrm{Fix}(\Lambda)=\{\rho \in \sh \, | \, \Lambda(\rho)=\rho\}$, where $\sh$ denotes the set of quantum states in the Hilbert space $\hi$, and $\hO$ denotes the set of all observables.
\begin{theorem}\label{thmstates}
    Let  $\hT \subset \sh$. Then there is an EBC $\Lambda$ such that $\hT \subset \mathrm{Fix}(\Lambda)$ if and only if $(\hT,\hO,\hO)$ is broadcastable.
\end{theorem}
The proof of this theorem, which is based on an averaging argument over repeated broadcasting channels, is presented in the Appendix~\ref{appendix:thmstates}. The following Corollary is a direct application of the no-broadcasting theorem \cite{Barnum2006}. 
 
 \begin{corollary}
     A set of quantum states $\hT$ does not allow for a proof of contextuality of quantum theory if and only if $\hT$ is commutative.
 \end{corollary}

The problem of identifying sets of measurements allowing for a proof of contextuality of quantum theory also reduces to a fixed point problem. One sees from Eq.~(\ref{ontologicalmodel3}) that a non-contextual model entails that $M_k=\sum_{\lambda} \tr{\sigma_{\lambda}M_k}G_{\lambda}$. In other words, the measurements need to be fixed points of the Heisenberg picture $\Lambda^*$ of an EBC $\Lambda$.

Before stating the result for measurements, we need to define \emph{instruments} and \emph{repeatability} of measurements. An instrument is a collection of trace-non-increasing completely positive linear maps $\{\mathcal I_k\}_k$ such that $\sum_k \mathcal I_k$ is a quantum channel. An instrument $\{\mathcal I_k\}_k$ is said to implement the measurement $\{A_k\}_k$ if $\mathcal I_k^*(I)=A_k$ for all $k$. If there exists an instrument $\{\mathcal I_k\}_k$ implementing $\{A_k\}_k$ such that $\mathcal I_k^*(A_k)=A_k$ for all $k$, then $\{A_k\}_k$ is called repeatable. In the following $\Vert \cdot \Vert$ denotes the operator norm, and classical post-processing of a POVM $\{G_\lambda\}_\lambda$ refers to a POVM $A_k=\sum_\lambda p(k|\lambda)G_\lambda$, where $p(\cdot|\lambda)$ is a probability distribution for each $\lambda$.
\begin{theorem}\label{thmmeasurement}
    For a set of observables $\hA \subset \hO$, the following statements are equivalent.
    \begin{enumerate}
        \item $\hA$ does not allow for a proof of contextuality of quantum theory.
        \item $(\sh,\hA,\hA)$ is broadcastable. 
        \item Every $\{A_k\}_k \in \hA$ is a classical post-processing of a single norm-1 POVM $\{G_\lambda\}_\lambda$, i.e.\ $\Vert G_{\lambda}\Vert=1$, $\forall \lambda$.
        \item Every $\{A_k\}_k \in \hA$ is a classical post-processing of a single repeatable measurement. 
    \end{enumerate}
\end{theorem}

The proof is presented in the Appendix~\ref{appendix:thmmeasurement}. We note that the equivalence between conditions $2$ and $3$ is the no-broadcasting theorem for measurements \cite{Heinosaari_2023}, which is here utilized to give more structure for measurements fulfilling condition $1$.
\par

To understand condition $3$ better, we investigate a few example cases. First, commutative sets of measurements fulfill the conditions of Theorem \ref{thmmeasurement}. This is due to a theorem by von Neumann \cite[Theorem 11.3]{busch16} stating that commutative self-adjoint operators are functions of a common self-adjoint operator.

Second, the conditions of Theorem \ref{thmmeasurement} do not imply commutativity, as there are non-commuting and broadcastable measurements \cite{Heinosaari_2023}. An explicit example is given by a norm-1 measurement. These are measurements that have a projective part in one subspace and possibly a non-projective part in the orthogonal complement. Intuitively, the deterministic or projective part allows one to pick the classical information $\lambda$ that a non-contextual model can pass through. More precisely, let $\hi$ be a 5-dimensional Hilbert space with the orthonormal basis $\{\ket{i}\}_{i=0}^{4}$. Define $\ket{+}$ and $\ket{-}$: $\ket{\pm}=\frac{1}{\sqrt{2}}(\ket{0}\pm \ket{1})$. Let $a \in (0,1)$ and define the following three-valued POVM:
\begin{align*}
E_1&:=\kb{2}{2}+a\kb{0}{0}+(1-a)\kb{+}{+} \\
E_2&:=\kb{3}{3}+a\kb{1}{1} \\
E_3&:=\kb{4}{4}+(1-a)\kb{-}{-}
\end{align*}
This POVM is clearly noncommutative, since for example $E_2E_3\neq E_3E_2$. 
Define then the EB-channel $\Lambda$ as
$\Lambda^*(A):=\sum_{z=1}^3 \tr{A \kb{z+1}{z+1}} E_z$.
Then we have 
\begin{align*}
\Lambda^*(E_k)=\sum_{z=1}^3 \tr{E_k \kb{z+1}{z+1}} E_z=\sum_{z=1}^3 \delta_{kz} E_z=E_k
\end{align*}
Therefore there exist measurements that are non-commutative but do not allow for a proof of contextuality of quantum theory. 

Finally, there are special sets of measurements for which the conditions of Theorem \ref{thmmeasurement} imply commutativity. One example is given by rank-1 measurements. As stated above, a POVM $\{G_\lambda\}_\lambda$ is of norm 1 if and only if $G_\lambda=P_\lambda+F_\lambda$ for all $\lambda$, where $\{P_\lambda\}_\lambda$ is a projection valued observable acting on a subspace (with $P_\lambda\ne 0$ for all $\lambda$) and $\{F_\lambda\}_\lambda$ a POVM acting on the orthogonally complemented subspace \cite{Pello7}. Moreover, if a rank-1 POVM $\{A_k\}_k$ is a postprocessing of a norm-1 $\{G_\lambda\}_\lambda$, then $F_\lambda=0$ for all $\lambda$ and $\{P_\lambda\}_\lambda$ is a rank-1 basis measurement, i.e.\ $P_\lambda=\kb{\fii_\lambda}{\fii_\lambda}$ for some orthonormal basis vectors $\fii_\lambda$ \cite{pellonpaa14}. Now $A_k=p_{k}P_{\lambda_k}$ where $p_{k}\in[0,1]$ so $\{A_k\}_k$ is commutative. This is summarized in the following Corollary.

\begin{corollary}\label{Corollary4}
    Suppose $\{A_k\}_k$ is a rank-1 POVM. Then it fulfills the equivalent conditions of Theorem \ref{thmmeasurement} if and only if it is commutative.
\end{corollary}
    
\section{Relation to other notions of classicality} Non-contextuality can be seen as a dividing line between classical and quantum behaviour, in that it asks whether a given theory is simplex embeddable \cite{Schmid2021}. In a simplex theory, states have a unique decomposition into extreme points. Our Theorem \ref{thmmeasurement} helps one to relate contextuality to other notions of non-classicality.

As the first example, we take joint measurability \cite{HeinosaariInvitation,GuhneJMReview}. In Ref.~\cite{TavakoliUola} it was shown that a set of measurements allows for a proof of preparation contextuality of quantum theory if and only if the set is not jointly measurable, i.e. not a post-processing of any single POVM. In contrast, Theorem \ref{thmmeasurement} requires such single POVM to be norm-1. This complements an example given in Ref.~\cite{Selby2023} of POVMs that allow for a proof of full contextuality, i.e. do not fulfill the conditions of Theorem \ref{thmmeasurement}, but are jointly measurable. There are indeed considerably more measurements allowing for a proof of full contextuality than those allowing for a proof of preparation contextuality: Norm-1 POVMs can not have more outcomes than the dimension of the underlying Hilbert space. Combining this with a dimension counting argument shows that any subset of measurements not allowing for a proof of full contextuality of quantum theory has zero volume. On the contrary, it is well known that the jointly measurable subset of measurements has a non-zero volume \cite{Reeb2013}.

As another example, the notion of inherent measurement disturbance goes back to Heisenberg's microscope, and has a clear operational formulation, cf. Refs.~\cite{Heinosaari10,BLWReview,busch16}. Here, we follow Ref.~\cite{Heinosaari10}: a measurement $\{A_a\}$ does not disturb a measurement $\{B_b\}$ if there is an instrument $\{\mathcal I_a\}$ implementing $\{A_a\}$ such that $\sum_a\mathcal I_a^*(B_b)=B_b$ for all outcomes $b$. In the qubit case, non-disturbance reduces to commutativity \cite{Heinosaari10}, but in qutrits and beyond there are non-commuting and non-disturbing measurements \cite{Heinosaari10}. An example is given by the following pair of binary qutrit POVMs \cite{Heinosaari10}:
$A_1=\frac{1}{4}\big(2|0\rangle\langle 0|+|2\rangle\langle 2|+\sqrt{2}(|0\rangle\langle 2|+|2\rangle\langle 0|)\big)$ and $B_1=\frac{1}{2}\big(2|0\rangle\langle 0|+|2\rangle\langle 2|\big)$, with $A_2=I-A_1$ and $B_2=I-B_1.$ Taking an optimal non-disturbing implementation $\{\mathcal I_a\}_a$ of $\{A_a\}_a$, we define the POVM $\{\mathcal I_a^*(B_b)\}_{a,b}$. This POVM does not fulfill the conditions of Theorem \ref{thmmeasurement}, as it is not a post-processing of a norm-1 measurement: If the pair given by $\{A_a\}$ and $\{B_b\}$ were a post-processing of a norm-1 POVM $\{G_{\lambda}\}$, the linear independence of $\{A_2,B_1,B_2\}$ would imply that $\dim \mathrm{lin}(\{G_{\lambda}\})\geq 3$. Since $G_{\lambda}=P_{\lambda}+F_{\lambda}$, where $\{P_{\lambda}\}$ is a projection valued measure, i.e. a POVM consisting of projections, this implies that $\dim \mathrm{lin}(\{G_{\lambda}\})= 3$ and furthermore that $G_{\lambda}=P_{\lambda}$, as the underlying Hilbert space is $\mathbb{C}^3$. This in turn would imply that $A$ and $B$ are classical post-processings of a common projection valued measure, i.e. commutative. This is a contradiction. Hence, operational non-contextuality in quantum theory is more restrictive than non-disturbance.

The above example can be directly applied to temporal correlations. It is well-known that a temporal correlation scenario consisting of non-disturbing measurements can be described by a macrorealistic hidden variable model for all input quantum states \cite{Uola2019macro}. These are models similar to Bell's local models with the locality assumption replaced by a non-invasiveness assumption. Therefore, the sequential POVM $\{\mathcal I_a^*(B_b)\}_{a,b}$ does allow for a proof of contextuality, but not for a violation of macrorealism.

\section{Pseudo-broadcasting and conditions for contextuality in a subtheory} One can also define operational contextuality for other theories than quantum. We are here interested in theories that consist of some convex subset of quantum states and a convex set of effects $E$, i.e. operators with $0 \leq E\leq I$, that includes the identity operator and yes-no questions, i.e., if an effect $E$ is part of the theory, then also $I-E$ is. Such a theory is called measurement and preparation non-contextual if
\begin{align}
    \tr{\rho E}=\sum_{\lambda} \tr{\rho \tilde G_{\lambda}}\tr{\tilde\sigma_{\lambda}E}
    \label{ontologicalmodel4}
\end{align}
for all states $\rho$ and effects $E$ of the theory. Here $\{\tilde G_\lambda\}_\lambda$ and $\{\tilde \sigma_\lambda\}_\lambda$ satisfy $\tr{\rho \tilde G_{\lambda}} \geq 0$ and $\sum_\lambda \tr{\rho \tilde G_{\lambda}} = 1$ for all states of the theory, and $\tr{\tilde\sigma_{\lambda} E} \geq 0$ and $\tr{\tilde\sigma_{\lambda}} = 1$ for all $\lambda$ and all effects $E$ of the theory. Note that this is different from full quantum theory, where positivity is required for all quantum states and measurements. Hence, non-contextuality in subtheories is a weaker notion.

One can also relax broadcasting by requiring that the broadcasting map $\Lambda$ in Eq.~(\ref{kakkonen}) and Eq.~(\ref{kolmone}) is not necessarily completely positive, but only by requiring that it preserves positivity of all probabilities that can be produced from $(\hT,\hA_1,\dots,\hA_n)$. That is, we say that $(\hT,\hA_1,\dots,\hA_n)$ is \btn{n}\emph{-pseudo-broadcastable} if there is a trace-preserving map $\Xi: \lh \to \hL(\hi^{\otimes n})$ such that for all $i \in \{1,\dots, n\}$, $\rho \in \hT$, and $\{A_{i,k}\}_k \in \hA_i$ it holds that
\begin{align}
\tr{A_{i,k} \tr_{i^c} \Xi(\rho)} &= \tr{A_{i,k}\rho}, \label{pseudobraodcasting-partialTrace} \\
\tr{\Xi(\rho) \otimes_{i=1}^n A_{i,k}} &\geq 0. \label{pseudobraodcasting-positivity}
\end{align}
Clearly \eqref{pseudobraodcasting-positivity} is a relaxation of the complete positivity of the broadcasting channel $\Lambda$ and one can find examples that are pseudo-broadcastable for all $n$ but not broadcastable, see the Appendix~\ref{appendix:sicpovm} for an example of a qubit symmetric informationally complete POVM having such property. We will be mostly interested in the case when $\hA_1 = \ldots = \hA_n$, we will denote the tuple as $(\hT, \hA)$, where $n$ will be understood from context.

The combined notion of measurement and preparation non-contextuality of a subtheory can be seen to be equivalent to \btn{n}-pseudobroadcasting for all $n \in \N$.
\begin{theorem}\label{thmgpt}
    A subtheory of quantum theory characterized by the allowed states $\hT$ and allowed measurements $\hA$ is measurement and preparation non-contextual if and only if $(\hT, \hA)$ is \btn{n}-pseudo-broadcastable for all $n \in \N$.
\end{theorem}
The proof, based on recent results in monogamy of ordered vector spaces and general probabilistic theories is relegated to the Appendix~\ref{appendix:thmgpt}.

\begin{figure}
\centering
\includegraphics[width=\linewidth]{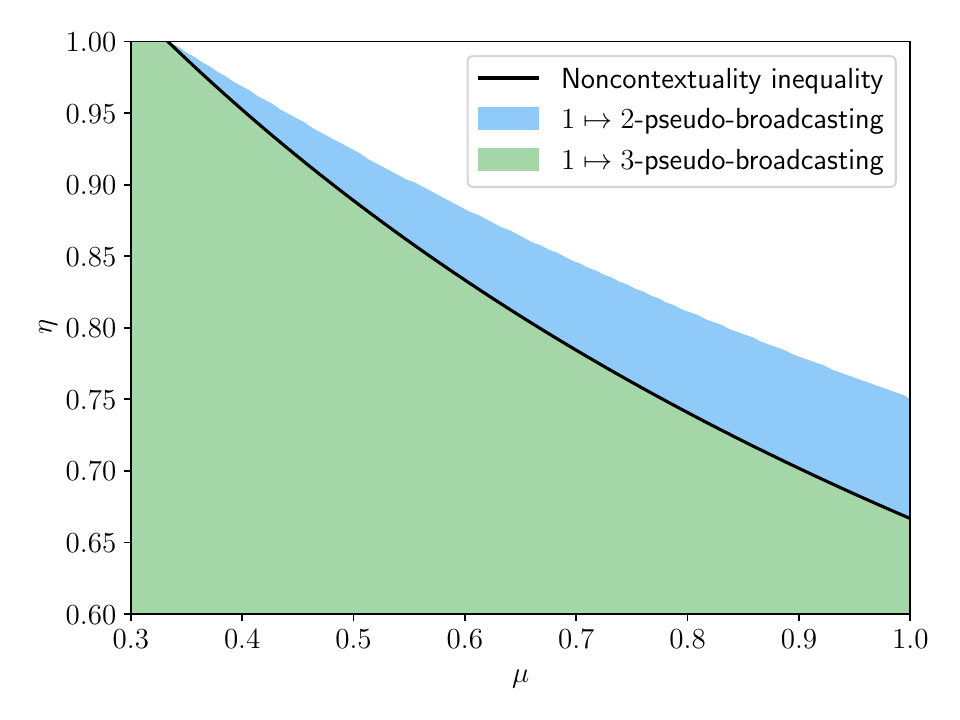}
\caption{The regions for pseudobroadcastability of the preparations and measurements considered in \cite{mazurek2016experimental}. $\mu$ and $\eta$ are unitless parameters of the dephasing channels acting on the preparations and depolarizing channel acting on the measurements, respectively. The green region where the scenario is \btn{3}-pseudobroadcastable coincides with the region where the inequality presented in \cite{mazurek2016experimental} is violated up to numerical precision of our calculations.}
\label{fig:mazurek}
\end{figure}

One can also use the result of Theorem~\ref{thmgpt} to prove non-contextuality of sets of preparations and measurements: for finite sets of preparations and measurements and for fixed $n \in \N$, checking whether there exists \btn{n}-pseudobroadcasting linear map is a semidefinite programming (SDP) problem since the conditions \eqref{pseudobraodcasting-partialTrace} and \eqref{pseudobraodcasting-positivity} are linear in the linear map $\Xi$. We use this to investigate contextuality of the preparations and measurements used in \cite{mazurek2016experimental}. These consist of six operators, all of which lie on the $x-z$ plane of the Bloch sphere and are separated by the angle $\pi/3$. We will moreover assume that a dephasing channel with parameter $\mu$ is acting on the preparations and a depolarizing channel with parameter $\eta$ is acting on the measurements, we will use $\mu$ and $\eta$ as parameters with respect to which we will investigate whether the scenario is contextual or not. The results are depicted in Fig.~\ref{fig:mazurek} and the details are relegated to the Appendix~\ref{appendix:mazurek}. For $n=2$ we obtain a region of the parameter space where the scenario is \btn{2}-pseudobroadcastable and thus, according to Theorem~\ref{thmgpt}, outside of this region the scenario must be contextual. For $n = 3$ we obtain a smaller region of the parameter space where the scenario is \btn{3}-pseudobroadcastable, this parameter space coincides up to numerical precision with the region where the inequality presented in \cite{mazurek2016experimental} is violated. This is likely not a coincidence since the scenario is generated by $3$ measurements and we conjecture that for a scenario generated by $k$ measurements \btn{k}-pseudobroadcasting coincides with non-contextuality. An intuition for why this conjecture may be true is given in the Appendix~\ref{appendix:thmgpt} after the proof of Theorem~\ref{thmgpt} as this intuition is based on the techniques used in the respective proof.

It is noteworthy that while our method agrees with the inequality presented in \cite{mazurek2016experimental}, our method can be easily applied to other similar scenarios and solved numerically without the need for finding additional contextuality inequalities. This example also shows a scenario which is \btn{2}-pseudobroadcastable but not \btn{3}-pseudobroadcastable, thus showing that unlike in the case of broadcasting, \btn{(n+1)}-pseudobroadcasting is not implied by \btn{n}-pseudobroadcasting.

\section{Conclusions}
By proving a one-to-one connection between non-contextuality and broadcasting in quantum theory, we have provided an operational, information theoretic characterization of the former. This not only shows that non-contextuality connects to another well-known concept in quantum information theory, but it also allows one to transfer insights between the two fields. As an example, there have been various efforts to understand the interplay between known quantum properties and proofs of contextuality \cite{liang11,yu13,Kunjwal14,kunjwal2014note,LostaglioWork,Xu19,TavakoliUola,plavala2022incompatibility,Selby2023,plavala2024contextuality}, but the characterization of full contextuality was missing until now. This was reached here for both quantum states and measurements by utilizing the no-broadcasting theorem.

We further showed that our results help one to relate non-contextuality to other notions of classicality, such as joint measurability, non-disturbance and Leggett-Garg macrorealism. Our results suggest that non-contextuality is the most restrictive notion of these four in quantum theory. This complements the recent results of Ref.~\cite{SchmidMacrorealism} on connections between macrorealism and non-contextuality, and shows that measurements not allowing for a proof of contextuality of quantum theory need to have measure zero.

Finally, we showed that our results are not restricted to quantum theory, but they also apply to subtheories by using the notion of pseudo-broadcasting. This notion can be easily decided numerically, hence, providing a method for finding witnesses for contextuality in subtheories.

Whether the results provided in this article carry over to continuous variable systems is left as an intriguing open question. The solution may rely on the used extension of non-contextuality to the continuous realm. One possible extension can be found by drawing inspiration from our results and known results on broadcasting in infinite-dimensional systems \cite{kuramochi2020}. However, it is beyond the scope of this work to investigate how this would relate to the existing approaches to contextuality in such systems \cite{Plastino2010,McKeown2011,Su2012,Asadian2015,LaversanneFinot2017,Barbosa2022,Booth2022}.

\begin{acknowledgments}
We are thankful to Arindam Mitra for discussions to Jef Pauwels and Isma\"{e}l Septembre for providing feedback on an earlier version of the manuscript. PJ and RU acknowledge support from the Swiss National Science Foundation (Ambizione PZ00P2-202179). MW is grateful for the support from the Swiss National Science Foundation (Ambizione PZ00P2\textunderscore208779).
MP acknowledges support from the Deutsche Forschungsgemeinschaft (DFG, German Research Foundation, project numbers 447948357 and 440958198), the Sino-German Center for Research Promotion (Project M-0294), the German Ministry of Education and Research (Project QuKuK, BMBF Grant No. 16KIS1618K), the DAAD, and the Alexander von Humboldt Foundation.
\end{acknowledgments}

\bibliography{bibbi}{}

\onecolumngrid
\appendix

\section{Proof of Theorem \ref{thmstates}} \label{appendix:thmstates}
($\Rightarrow$) Suppose that $\hT \subset \mathrm{Fix}(\Lambda)$ for some EB-channel $\Lambda$ with $\Lambda(\rho):=\sum_\lambda \tr{\rho G_\lambda} \sigma_\lambda$ for all $\rho$. Then one can define the joint channel $\Phi(\rho):=\sum_\lambda \tr{\rho G_\lambda} \sigma_\lambda \otimes \sigma_\lambda$ such that $\mathrm{tr}_i[\Phi(\rho)]=\rho$, $i=1,2$ for all $\rho \in \mathrm{Fix}(\Lambda)$, i.e.\ the broadcasting condition holds. \\

($\Leftarrow$) Suppose $(\hT,\hO,\hO)$ is broadcastable with the channel $\Lambda$. Then $\hT$ is $1\to n$-broadcastable for all $n\geq 2$. This can be seen by induction as follows. Let $\Lambda_2=\Lambda$ and assume that $\hT$ is $1 \to n$-broadcastable with the channel $\Lambda_n:\th \to \hT(\hi^{\otimes n})$. Define then the channels $i_n^{*}:\lh \to \hL(\hi^{\otimes n-1})$, $i_n^*(A):= A \otimes \left(\bigotimes_{k=1}^{n-2} I\right)$ and $\Lambda_{n+1}:=(\Lambda_n \otimes i_n) \circ \Lambda_n$. Now if $\rho \in \hT$, we have the following for all $A \in \lh$ and $k \in \{1,\dots,n\}$, where $k^c$ denotes the complement of the set $\{k\}$:
     \begin{align*}
     &\tr{A\mathrm{tr}_{k^c}[\Lambda_{n+1}(\rho)]} \\ &=\tr{(\Lambda_n^*(I \otimes \cdots \otimes A \otimes \cdots \otimes I) \otimes I \otimes \cdots \otimes I) \Lambda_n(\rho) }\\
     &=\tr{\Lambda_n^*( I \otimes \cdots \otimes A \otimes \cdots \otimes I) \rho }=\tr{A \rho}.
     \end{align*}
     If $k=n+1$, then 
     \begin{align*}
     \tr{A\mathrm{tr}_{
     (n+1)^c}[\Lambda_{n+1}(\rho)]}&=\tr{(I \otimes  A \otimes  I \otimes \cdots \otimes I) \Lambda_n(\rho) }\\
     &=\tr{A \mathrm{tr}_{2^c}[\Lambda_n(\rho)]}=\tr{A \rho}.
     \end{align*}
Thus the set $\hT$ is $1 \to n+1$-broadcastable. By induction we see that $\hT$ is $1 \to n$-broadcastable for all $n \in \N$. \par 
Let us denote the $1 \to n$ broadcasting channel still by $\Lambda_n$. Furthermore, let $S_n$ be the $n$:th symmetric group with the unitary representation $\pi \mapsto U_{\pi}$ where $U_{\pi}:\hi^{\otimes n} \to \hi^{\otimes n}$ with $U_{\pi}\left( \bigotimes_{i=1}^n \fii_i\right)=\bigotimes_{i=1}^n \fii_{\pi(i)}$. Then for all $n \in \N$ we define the channel $\tilde{\Lambda}_n:\th \to \hT(\hi^{\otimes n})$ by
    \begin{align*}
    \tilde{\Lambda}_n (\rho)=\frac{1}{n!}\sum_{\pi \in S_n} U_{\pi} \Lambda_n(\rho) U_{\pi}^*.
    \end{align*}
Let then $k \in \{1,\dots, n\}$. Now 
     \begin{align*}
     \mathrm{tr}_{k^c}[\tilde{\Lambda}_n(\rho)]&=\frac{1}{n!}\sum_{\pi \in S_n} \mathrm{tr}_{k^c} [U_{\pi} \Lambda_n(\rho) U_{\pi}^*] \\
     &=\frac{1}{n!}\sum_{\pi \in S_n} \mathrm{tr}_{\pi(k)^c} [\Lambda_n(\rho)]\\
     &=\frac{1}{n}\sum_{k=1}^n \mathrm{tr}_{k^c}[\Lambda_n(\rho)]=:\Phi_n(\rho).
     \end{align*}
Here if $\rho \in \hT$, then it is a fixed point of $\Phi_n$ by the $1 \to n$-broadcasting condition. Furthermore, $\Phi_n$ is $n$-self compatible with the joint channel $\tilde{\Lambda}_n$. Now since the set of quantum channels is compact in the diamond norm, we have that there is a convergent subsequence $(\Phi_{n_m})_m$ of $(\Phi_n)_n$. Furthermore each $\Phi_n$ is especially the first marginal of the symmetric broadcast channel $\tilde{\Lambda}_n$. Therefore by \cite[Theorem 5]{chiribella2011} we see that for every $n$ there exists an entanglement breaking channel $\Lambda_n^{EB}:\th \to \th$ such that 
    \begin{align*}
     \Vert \Phi_n - \Lambda_n^{EB} \Vert_{\Diamond} \leq \frac{2d^2}{n}.
     \end{align*}
Here $d=\dim \hi$ and $\Vert \cdot \Vert_{\Diamond}$ denotes the diamond norm. Now as the set of entanglement breaking channels is also compact in the diamond norm, we have that the sequence $(\Lambda_{n_m}^{EB})_m$ has a convergent subsequence $(\Lambda_{n_{m_j}}^{EB})_{j}$. Let the limit of $(\Lambda_{n_{m_j}}^{EB})_{j}$ be $\Lambda^{EB}$ and the limit of $(\Phi_{n_m})_m$ be $\Phi$. Now
\begin{align*}
&\Vert \Phi -\Lambda^{EB} \Vert_{\Diamond} \\ &\leq \Vert \Phi-\Phi_{n_{m_j}} \Vert_{\Diamond} +\Vert \Phi_{n_{m_j}} -\Lambda_{n_{m_j}}^{EB}\Vert_{\Diamond}+\Vert \Lambda_{n_{m_j}}^{EB}-\Lambda^{EB}\Vert_{\Diamond} \\
&\leq \Vert \Phi-\Phi_{n_{m_j}} \Vert_{\Diamond} +\frac{2d^2}{n_{m_j}}+\Vert \Lambda_{n_{m_j}}^{EB}-\Lambda^{EB}\Vert_{\Diamond}.
\end{align*}
Letting $j \to \infty$ we see that $\Phi=\Lambda^{EB}$. \par 
Finally we show that $\hT \subset \mathrm{Fix}(\Phi)$. Let $\rho \in \hT$. Now 
\begin{align*}
\Vert \Phi(\rho)-\rho \Vert_1 = \Vert \Phi(\rho)-\Phi_{n_m}(\rho)\Vert_1 \to 0,
\end{align*}
as $m \to \infty$. Therefore $\rho$ is a fixed point of the entanglement breaking channel $\Phi$.

\section{Proof of Theorem \ref{thmmeasurement}} \label{appendix:thmmeasurement}
The equivalence $2 \Leftrightarrow 3$ is Theorem 2 of \cite{Heinosaari_2023} and $3 \Leftrightarrow 4$ is a known equivalence \cite{busch96}. Let us prove the implication $1\Rightarrow 2$. Let $\{A_k\}_k \in \hA$. Then all $A_k$ are fixed points of a Heisenberg picture EB-channel $\Lambda^*$ with $\Lambda^*(A)=\sum_{\lambda} \tr{\sigma_{\lambda}A} G_{\lambda}$. Here $\{\sigma_{\lambda}\}_{\lambda}$ is a family of states and $G$ a POVM. Define then the channel $\Phi(\rho)=\sum_{\lambda} \tr{G_{\lambda}\rho} \sigma_{\lambda} \otimes \sigma_{\lambda}$. Then we have that $\tr{A_k \tr_i{\Phi(\rho)}}=\tr{A_k \Lambda(\rho)}=\tr{A_k \rho}$ for all $\rho \in \sh$ and $i=1,2$. Thus the broadcasting condition holds. \par 
Finally we need to prove the implication $3 \Rightarrow 1$. This can be seen very directly from the sufficiency part of the proof of Theorem 2 in \cite{Heinosaari_2023} as the marginal channels in the proof are obviously entanglement breaking. For completeness, the argument is as follows. Suppose that $E$ is a post-processing of a norm-1 POVM $G$. Then for every $G_\lambda$ there is a unit vector $\ket{\lambda}$ such that $G_\lambda \ket{\lambda}=\ket{\lambda}$. Since $\sum_{\lambda'}\ip{\lambda'}{G_\lambda|\lambda'}=1$, we see that $\ip{\lambda'}{G_\lambda|\lambda'}=\delta_{\lambda\lambda'}$. Therefore we define the EB-channel 
\begin{align*}
\Lambda^*(A):=\sum_{\lambda'}\ip{\lambda'}{A|\lambda'} G_{\lambda'}
\end{align*}
Since $\ip{\lambda'}{G_\lambda|\lambda'}=\delta_{\lambda\lambda'}$, all $G_\lambda$ are fixed points by construction. Therefore any post-processing of the POVM $G$ is also a fixed point by linearity.

\section{There exist tuples $(\hT,\hA)$ that are not broadcastable, but are $1 \mapsto n$-pseudo-broadcastable} \label{appendix:sicpovm}
Let $\{\ket{i}\}_{i=0,1}$ be an orthonormal basis for the qubit Hilbert space $\mathcal{H} = \mathbb{C}^2$ and let $\{E_i\}_{i=1}^4$ be the symmetric, informationally complete POVM (SIC-POVM) in qubit. In other words $E_i=\frac{1}{2}\kb{\fii_i}{\fii_i}$ with 
\begin{align}
\ket{\fii_1}&:=\ket{0} \\
\ket{\fii_2}&:=\frac{1}{\sqrt{3}}\ket{0}+\sqrt{\frac{2}{3}}\ket{1} \\
\ket{\fii_3}&:=\frac{1}{\sqrt{3}}\ket{0}+\sqrt{\frac{2}{3}}e^{i \frac{2\pi}{3}}\ket{1} \\
\ket{\fii_4}&:=\frac{1}{\sqrt{3}}\ket{0}+\sqrt{\frac{2}{3}}e^{i \frac{4\pi}{3}}\ket{1}. 
\end{align}
These are indeed symmetric in the sense that their Hilbert-Schmidt inner product is of special form: $\tr{\kb{\fii_i}{\fii_i}\kb{\fii_j}{\fii_j}}=\frac{2\delta_{ij}+1}{3}$. \par 
We will now show that the single POVM theory $(\sh,\{\{E_i\}_{i=1}^4\})$ is not broadcastable, but is $1 \mapsto n$ pseudo-brodacastable for all $n \in \N$. The fact that $(\sh,\{\{E_i\}_{i=1}^4\})$ is not broadcastable follows immediately from the fact there is no universal broadcasting. This is since $\mathrm{lin}\{E_1,E_2,E_3,E_4\}=\hL(\mathbb{C}^2)$ by informational completeness. \par 
Let us then show that $(\sh,\{\{E_i\}_{i=1}^4\})$ is $1 \mapsto n$ pseudo-broadcastable. For this, we define the $1 \mapsto n$ pseudo-broadcasting map $\Xi_n:\hL(\mathbb{C}^2) \to \hL((\mathbb{C}^2)^{\otimes n})$ as follows. For all $\rho \in \hL(\mathbb{C}^2)$ we let
\begin{align}
\Xi_n(\rho):=\sum_{i=1}^4 \tr{E_i \rho}(3 \kb{\fii_i}{\fii_i}-I)^{\otimes n}.
\end{align}
This is obviously trace-preserving and indeed fulfills the conditions of a valid $1 \mapsto n$ pseudo-broadcasting map, which can be seen as follows.
\begin{align}
&\tr{(E_{i_1}\otimes E_{i_2} \otimes \cdots E_{i_n})\Xi_n(\rho)} \\
&=\frac{1}{2^n}\sum_{j=1}^4 \tr{\rho E_j} \prod_{k=1}^n (3\tr{\kb{\fii_j}{\fii_j}\kb{\fii_{i_k}}{\fii_{i_k}}}-1) \\
&=\frac{1}{2^n}\sum_{j=1}^4 \tr{\rho E_j} \prod_{k=1}^n  2 \delta_{i_k j} \geq 0
\end{align}
Therefore positivity holds. Furthemore, by the equation above we also get 
\begin{align}
&\tr{(I\otimes I \otimes \cdots \otimes E_{i_j} \otimes \cdots I)\Xi_n(\rho)} \\
&=\sum_{i_1\dots i_{j-1}i_{j+1}\dots i_n}\tr{(E_{i_1}\otimes E_{i_2} \otimes \cdots E_{i_n})\Xi_n(\rho)} \\
&=\frac{1}{2^n}\sum_{k=1}^4 \tr{\rho E_k}   2^n \delta_{i_j k} =\tr{\rho E_{i_j}}
\end{align}
Thus the pseudo-broadcasting condition holds. Since $n \in \N$ was arbitrary, $(\sh,\{\{E_i\}_{i=1}^4\})$ is $1 \mapsto n$ pseudo-broadcastable for all $n \in \N$.

\section{Proof of Theorem~\ref{thmgpt}} \label{appendix:thmgpt}
If $(\hT, \hA)$ is non-contextual, then there are $G_\lambda$ and $\sigma_\lambda$ such that for any $\rho \in \hT$ and $M \in \hA$ we have $\tr{\rho M} = \sum_\lambda \tr{\rho G_\lambda} \tr{\sigma_\lambda M}$, note that we can, without loss of generality, choose the operators $G_\lambda$ and $\sigma_\lambda$ such that $\tr{\sigma_\lambda} = 1$. Let $n \in \N$, we now construct $\Xi_n: \lh \to \hL(\hi^{\otimes n})$ as
\begin{equation}
     \Xi_n(\rho) = \sum_\lambda \tr{\rho G_\lambda} \sigma_\lambda^{\otimes n}.
\end{equation}
Since $\tr{\sigma_\lambda} = 1$ we have
\begin{equation}
     \tr{\tr_{i^c}[\Xi_n(\rho)] M} = \sum_\lambda \tr{\rho G_\lambda} \tr{\sigma_\lambda M} = \tr{\rho M}
\end{equation}
for all $\rho \in \hT$ and $M \in \hA$. Moreover
\begin{equation}
     \tr{\Xi_n(\rho) \otimes_{i=1}^n M_i} = \sum_\lambda \tr{\rho G_\lambda} \prod_{i=1}^n \tr{\sigma_\lambda M_i} \geq 0
\end{equation}
since $\tr{\sigma_\lambda M_i} \geq 0$ for all $M_1, \ldots, M_n \in \hA$ and $\tr{\rho G_\lambda} \geq 0$. Thus $\Xi_n$ is the \btn{n}-pseudo-broadcasting map for $(\hT, \hA)$.

Now assume that $(\hT, \hA)$ is \btn{n}-pseudo-bradcastable for all $n \in \N$. We will use $\hS(\hA)$ to denote the state space over $\hA$, that is $\hS(\hA)$ is the set of linear functionals on $\linspan(\hA)$ such that for $\varphi \in \hS(\hT)$ we have $\varphi(M) \geq 0$ for all $M \in \hA$ and $\varphi(I) = 1$. Note that to every $\varphi \in \hS(\hT)$ one can find at least one operator $\sigma \in \lh$ such that $\varphi(M) = \tr{\sigma M}$ since $\linspan(\hA) \subset \lh$. We will also need to define maximal tensor product of the state spaces $\hS(\hA)$: we define $\hS(\hA) \tmax \hS(\hA) = \hS(\hA)^{\tmax 2}$ to be the set of linear functionals on $\linspan(\hA)^{\otimes 2}$ such that for $\varphi_2 \in \hS(\hT)^{\tmax 2}$ we have $\varphi(M_1 \otimes M_2) \geq 0$ for all $M_1, M_2 \in \hA$ and $\varphi(I \otimes I) = 1$. Analogically, we define $\hS(\hA)^{\tmax k}$ for $k \in \N$ to be the set linear functionals on $\linspan(\hA)^{\otimes k}$ such that for $\varphi_k \in \hS(\hT)^{\tmax k}$ we have $\varphi(\otimes_{i=1}^k M_i) \geq 0$ for all $M_1, \ldots, M_k \in \hA$ and $\varphi(I^{\otimes k}) = 1$.

Let $\Xi_1: \hT \to \hS(\hA)$ be the linear map defined as $(\Xi_1(\rho))(M) = \tr{\rho M}$ for all $\rho \in \hT$ and $M \in \hA$. We will now argue that for every $n \in \N$ the map $\Xi_1$ has a $n$-copy extension, that is, for every $n \in \N$ there is a map $\Psi_n: \hT \to \hS(\hT)^{\tmax n}$ such that for any $\rho \in \hT$ and $M \in \hA$ we have
\begin{equation}
     (\Psi_n(\rho))(M \otimes I^{\otimes (n-1)}) = (\Psi_n(\rho))(I \otimes M \otimes I^{\otimes (n-2)}) = \ldots = (\Psi_n(\rho))(I^{\otimes (n-1)} \otimes M) = (\Xi_1(\rho))(M) = \tr{\rho M}.
\end{equation}
The $n$-copy extension of $\Xi_1$ are the \btn{n}-pseudo-broadcasting maps $\Xi_n$, that is, we can take $\Psi_n = \Xi_n$: the equation above corresponds to the conditions $\tr{M \tr_{i^c} \Xi_n(\rho)} = \tr{\rho M}$ for all $i \in \{1,\dots, n\}$ and $\Psi_n(\rho) \in \hS(\hA)^{\tmax n}$ is equivalent to $\tr{\Xi_n(\rho) \otimes_{i=1}^n M_i} \geq 0$ for all $M_1, \ldots, M_n \in \hA$ by definition. It now follows from the identification of linear maps with elements of the appropriate tensor product \cite[Proposition 6.9]{plavala2023general} that existence of $n$-copy extensions for all $n \in \N$ implies that the map $\Xi_1$ is separable \cite{aubrun2022monogamy}, i.e., measure-and-prepare. That is, there are operators $G_\lambda$ such that $\tr{\rho G_\lambda} \geq 0$ for all $\rho \in \hT$ and all $\lambda$ and $\varphi_\lambda \in \hS(\hA)$ such that
\begin{equation}
     \Xi_1(\rho) = \sum_\lambda \tr{\rho G_\lambda} \varphi_\lambda.
\end{equation}
By taking any of the operators $\sigma_\lambda$ corresponding to the functionals $\varphi_\lambda$ via $\varphi_\lambda(M) = \tr{\sigma_\lambda M}$ for any $M \in \hA$, we get
\begin{equation}
     \tr{\rho M} = (\Xi_1(\rho))(M) = \sum_\lambda \tr{\rho G_\lambda} \varphi_\lambda(M) = \sum_\lambda \tr{\rho G_\lambda} \tr{\sigma_\lambda M}.
\end{equation}

It was proved in \cite{aubrun2022monogamy} that if $\hS(\hA)$ is a Cartesian product of $k$ simplexes, then one only needs the existence of $n$-copy extensions for $n \leq k$ to prove that $\Xi_1$ is separable and thus measure-and-prepare. One can also see that $\hS(\hA)$ is a Cartesian product of $k$ simplexes if $\hA$ is generated by $k$ independent measurements. This leads to a conjecture that if $\hA$ is generated by $k$ measurements, then \btn{k}-pseudobroadcasting is equivalent to non-contextuality. This is only a conjecture since even if $\hA$ is generated by $k$ measurements, $\hS(\hA)$ in general is only a subset of a Cartesian product of $k$ simplexes. We leave resolving this issue and proving or disproving the conjecture for future work.

\section{\btn{n}-pseudobroadcasting in the scenario investigated in \cite{mazurek2016experimental}} \label{appendix:mazurek}
Consider the qubit Hilbert space, $\dim(\cH) = 2$. Let $t \in \{1,2,3\}$ and $b \in \{0,1\}$ then we will consider a scenario with preparations $\sigma_{t,b}$ and measurement effects $M_{t,b}$ given as

\begin{align}
&\sigma_{1,0} = M_{1,0} = \dfrac{1}{2} \left(I + \pZ \right),
&&\sigma_{1,1} = M_{1,1} = \dfrac{1}{2} \left(I - \pZ \right), \\
&\sigma_{2,0} = M_{2,0} = \dfrac{1}{2} \left(I + \dfrac{\sqrt{3}}{2} \pX - \dfrac{1}{2} \pZ \right),
 &&\sigma_{2,1} = M_{2,1} = \dfrac{1}{2} \left(I - \dfrac{\sqrt{3}}{2} \pX + \dfrac{1}{2} \pZ \right), \\
 &\sigma_{3,0} = M_{3,0} = \dfrac{1}{2} \left(I - \dfrac{\sqrt{3}}{2} \pX - \dfrac{1}{2} \pZ \right),
 &&\sigma_{3,1} = M_{3,1} = \dfrac{1}{2} \left(I + \dfrac{\sqrt{3}}{2} \pX + \dfrac{1}{2} \pZ \right),
 \end{align}
 where $\pX$, $\pZ$ are the usual Pauli operators. We will consider the dephasing channel acting on the preparations,
 \begin{equation}
 P_\mu(\rho) = \mu \rho + (1-\mu) \sum_{0}^1 \bra{i} \rho \ket{i} \ket{i}\bra{i},
 \end{equation}
 where $\ket{0}, \ket{1}$ is the standard computational basis of eigenvectors of $\pZ$, and the depolarizing channel acting on the measurement effects
 \begin{equation}
 D_\eta(E) = \eta E + (1-\eta) \dfrac{I}{2}.
 \end{equation}
 Our task is to determine for which pair of parameters $\mu, \eta$ are the preparations $P_\mu(\sigma_{t,b})$a and measurement effects $D_\eta(M_{t,b})$ contextual. Using the same approach as in \cite{mazurek2016experimental}, the scenario is contextual if the non-contextuality inequality presented in \cite{mazurek2016experimental} is violated, the non-contextuality inequality reads
 \begin{equation}
 \sum_{t = 1}^3 \sum_{b = 0}^1 \tr(\sigma_{t,b} M_{t,b}) \leq 5.
 \end{equation}
 By explicit calculations, one can show that this inequality is violated when
 \begin{equation}
 \eta \geq \dfrac{4}{3(1 + \mu)}.
 \end{equation}

 Another options is to check whether the scenario is \btn{2}-pseudobraodcastable for given $\mu, \eta$. This can be done via numerically solving the following semidefinite program (SDP):
 \begin{equation}
 \begin{split}
 \text{find} \quad &\Xi \in \hL(\cH^{\otimes 3}) \\
 \text{such that} \quad &\tr(\Xi [P_\mu(\sigma_{t,b}) \otimes D_\eta(M_{t',b'}) \otimes D_\eta(M_{t'',b''})]) \geq 0 \\
 &\tr(\Xi [P_\mu(\sigma_{t,b}) \otimes D_\eta(M_{t',b'}) \otimes I]) = \tr[P_\mu(\sigma_{t,b})  D_\eta(M_{t',b'})] \\
 &\tr(\Xi [P_\mu(\sigma_{t,b}) \otimes I \otimes D_\eta(M_{t',b'})]) = \tr[P_\mu(\sigma_{t,b})  D_\eta(M_{t',b'})] \\
 \text{for all} \quad &t, t', t'' \in \{1,2,3\}, b, b', b'' \in \{0,1\}
 \end{split} \label{12pseudobraodcastingSDP}
 \end{equation}
If the SDP \eqref{12pseudobraodcastingSDP} is not feasible, then the scenario is not \btn{2}-pseudobraodcastable and thus contextual. In a similar fashion one gets the SDP for \btn{3}-pseudobraodcasting, this reads
 \begin{equation}
\begin{split}
 \text{find} \quad &\Xi \in \hL(\cH^{\otimes 4}) \\
 \text{such that} \quad &\tr(\Xi [P_\mu(\sigma_{t,b}) \otimes D_\eta(M_{t',b'}) \otimes D_\eta(M_{t'',b''}) \otimes D_\eta(M_{t''',b'''})]) \geq 0 \\
 &\tr(\Xi [P_\mu(\sigma_{t,b}) \otimes D_\eta(M_{t',b'}) \otimes I \otimes I]) = \tr[P_\mu(\sigma_{t,b})  D_\eta(M_{t',b'})] \\
 &\tr(\Xi [P_\mu(\sigma_{t,b}) \otimes I \otimes D_\eta(M_{t',b'}) \otimes I]) = \tr[P_\mu(\sigma_{t,b})  D_\eta(M_{t',b'})] \\
 &\tr(\Xi [P_\mu(\sigma_{t,b}) \otimes I \otimes I \otimes D_\eta(M_{t',b'})]) = \tr[P_\mu(\sigma_{t,b})  D_\eta(M_{t',b'})] \\
 \text{for all} \quad &t, t', t'', t''' \in \{1,2,3\}, b, b', b'', b''' \in \{0,1\}
 \end{split}
 \end{equation}

\end{document}